\documentclass[12pt,english]{iopart}
\usepackage[T1]{fontenc}
\usepackage[latin1]{inputenc}
\usepackage{iopams}
\usepackage{babel}
\usepackage{psfrag}
\usepackage{graphicx}

\def\e{\varepsilon}

\begin{document}
\begin{flushright} {\footnotesize  LMU-ASC 36/05 \\ MPP-2005-37} 
\end{flushright}
\vspace{5mm}
\vspace{0.5cm}

\title{B-Inflation}

\author{Alexey Anisimov \footnote{anisimov@mppmu.mpg.de}}
\address{Max-Planck-Institut f\"ur Physik, Fohringer Ring 6,
D-80805, Munich, Germany\\
and \\
Institute of Theoretical and Experimental Physics,
117218, B.Cheremushkinskaya 25, Moscow, Russia}
\author{Eugeny Babichev \footnote{babichev@mppmu.mpg.de}}
\address{Max-Planck-Institut f\"ur Physik, Fohringer Ring 6,
D-80805, Munich, Germany\\
and \\
Institute for Nuclear Research of the Russian Academy of
Sciences, 60th October Anniversary Prospect 7a, 117312
Moscow, Russia}
\author{Alexander Vikman \footnote{vikman@theorie.physik.uni-muenchen.de}}
\address{Arnold-Sommerfeld-Center for Theoretical Physics, Department f\"ur Physik, 
 Ludwig-Maximilians-Universit\"at M\"unchen,
Theresienstr. 37, D-80333, Munich, Germany }

\pacs{98.80.Cq}

\begin{abstract}
{\small We propose a novel model of inflation based on a
large class of covariant effective actions containing only the second 
derivatives of a scalar field. The initial conditions leading to the 
inflationary solutions  in this  model are rather generic. The graceful exit 
from the inflationary regime is natural once the first order derivative terms are 
included. 
}{\normalsize \par}
\end{abstract}
\maketitle

\section{Introduction}

Inflationary paradigm has dominated cosmology since long ago and recently 
its predictions have attained excellent observational evidence. 
However at present there exists no preferred concrete inflationary 
scenario based on a convincing realistic high energy physics model. 
It is therefore interesting to explore every
single theoretical possibility which leads to inflation along with specific
predictions.
  
In the standard chaotic inflation scenario \cite{linde1} the massive 
inflaton field rolls down its potential. If its initial amplitude is much 
larger than Planck mass  $M_{Pl}$, the Hubble friction is 
large and inflaton field is in the 
"slow-roll" regime where its potential energy dominates over its kinetic
energy. In this case the energy density during "slow-roll" is nearly 
a constant  and the equation of state is nearly that of the cosmological 
constant. The exit from such inflationary regime occurs when the Hubble parameter
becomes of order of inflaton mass. When this happens the inflaton field starts to
behave as ordinary matter and later decays into light particles.
This simple model describes current observations rather well. There are 
numerous examples which develop this idea further \cite{pot_infl}. 
The predictions of all these models are quite similar, so that even the recent
accurate measurements \cite{wmap} did not put too much pressure on 
the viability of most of these models. Probably the most serious constraint is
the presently unobserved and, thus, necessarily small fraction of isocurvature 
fluctuations in the primordial spectrum. These fluctuations generically 
arise in  multi-field models. The other constraint comes from the 
nongaussianity in the primordial spectrum \cite{nong}. Future experiments on 
the detection of the gravitational background radiation \cite{gr}
left after inflation may impose further significant constraints on the 
inflationary models \cite{gr1}.  

Most of the inflationary models  can be classified as 
{\it potential induced}, i.e. inflation is driven by the potential terms in 
the Lagrangian of a scalar field. 
This is, however, not the only theoretical approach to implement 
the inflation. There is a distinct possibility that the inflationary 
regime was induced by the kinetic energy. The kinetic terms, of course, 
have to be of a nonstandard form, since for models with standard kinetic terms 
and without potential the universe is 
always decelerating.  
It is, however, possible to implement an inflationary evolution of the early 
Universe if one assumes 
a nonstandard  Lagrangian. A novel way to generate an inflationary 
regime using the above approach, has been proposed in \cite{k_inf} and 
dubbed  "{\it $k$-inflation}". The Lagrangian in this model was 
taken in the form of expansion in orders of the standard kinetic term 
$X={1\over 2}\left(\partial_{\mu}\phi\right)^2$:
\begin{equation}
{\mathcal L}=K(\phi)X+L(\phi)X^2+...=p(\phi,X).
\label{mukh}
\end{equation}
The system described by the  Lagrangian of this form was shown to 
dynamically enter the inflationary 
regime starting from rather generic initial conditions. There is an analog 
of a "slow-roll" regime, where the systems has an 
adiabatically changing, nearly de Sitter equation of state. 
The model possesses the energy-momentum tensor of the same structure
as for the perfect fluid with the energy density $\e$ and the pressure $p$.
Therefore the hydro-dynamical intuition
is useful for the analysis. 
At the end of {\it $k$-inflation} the equation 
of state $w=p/\varepsilon$ changes dynamically from $w\simeq-1$ to 
the  "{\it ultra-hard}" equation of state $w=1$ . Thus, exit from {\it $k$-inflation} 
occurs in a natural graceful manner. 


In this paper we propose a new model for inflation where the Lagrangian depends
only on the second derivatives of a scalar field.
The simplest scalar that can be built from a scalar field $\phi$ and 
its second covariant derivatives is the D'Alembert operator acting on $\phi$. 
We will restrict ourselves to this case. 
In fact, a large class of Lagrangians 
of the form
\begin{equation}
{\mathcal L}=Q(\Box \phi),
\label{firstLagr}
\end{equation}    
where $Q(\Box \phi)$ is a convex function, 
will be shown to drive an inflationary evolution of the universe 
from rather generic initial conditions. This theory has to be viewed as an
effective field theory where the first derivative terms are suppressed. One can, for
example, suppose that in the Lagrangian (\ref{mukh}) the functions $K(\phi)$ and 
$L(\phi)$ vanish in some limiting cases. Then the higher order derivative 
terms may dominate the dynamics of the system for some period of its evolution.  
The Lagrangians of this type were also considered in \cite{sec,Zhang} in 
various contexts\footnote{Note that, despite of  similarities 
with the model from Ref. \cite{Zhang}, our system does not have any potential and 
cannot be reduced to the noninteracting two field model \cite{Quintom} 
as it was done in the reference above. 
Thus the physical content of our model is completely 
different from that of Ref. \cite{Zhang}.}.

We will dub this model analogously to {\it $k$-inflation} as 
"{\it Box-inflation}" or for simplicity "{\it B-inflation}". 
The system described by this "truncated" Lagrangian does not have an exit from 
inflation, so after discussing the properties of this system during the inflationary 
stage we propose 
simple modifications of the model which provide a graceful exit from {\it B-inflation}.
One natural way to implement  this exit is to revive the kinetic terms which 
depend on $\phi$ and $X$. 
For the proposed modifications, the system will generically exit to the state with $w\simeq0$.
This is similar to the graceful exit in the case of chaotic inflation. 
The perturbation theory for the {\it B-inflation} as well as a detailed analysis of 
the observational consequences will be developed elsewhere \cite{pert}. 

\section{The Model}

We will first consider the action containing the D'Alembert 
operator $\Box \equiv g_{\mu\nu}\nabla^{\mu}\nabla^{\nu}$ (where $\nabla^{\mu}$ is the 
covariant derivative)  
acting on a scalar field $\phi$ minimally coupled to gravity:
\begin{equation}
  S \equiv S_g+S_{\phi} = \int d^4x\sqrt{-g}\left[-M_{Pl}^2 \frac{R}{2}+ 
    M^2 M_{Pl}^2\, Q\left(\frac{\Box \phi}{M^2 M_{Pl}}\right)\right].
  \label{action0}
\end{equation}
Here $Q$ is an arbitrary convex function ($Q''\geq const>0$, later we will see
that this inequality is important). 
We will only consider the case when  $Q(0)=0$ in order not to have the 
cosmological constant being inserted by hand. This underlines the fact that 
the inflationary regime in our model is a result of dynamics of the system.
In addition it is natural to assume that the system is symmetric with respect 
to the $Z_{2}$ transformations
$\phi\rightarrow-\phi$.
Thus we can take only even functions $Q$ into consideration.
The scale $M$, at this point, is an arbitrary parameter. 
For the purposes of this paper its value is not very important. 
We will leave the question of the value of 
this scale for a future discussion.
The ``strange'' choice of combination of masses for 
the $Q$--term is due to the following reason. Writing the 
action $S_\phi$ as a Taylor series in ($\Box \phi$) we find: 
\begin{equation}
  \label{TS}
  S_\phi=\int d^4x\sqrt{-g}\left[\frac{Q''(0)}{2 M^2}(\Box\phi)^2 + 
    \frac{Q^{(4)}(0)}{4!}\frac{(\Box\phi)^4}{M_{Pl}^2 M^6}+...\right].
\end{equation}
Now we can see that the initial choice of combination of 
masses leads to a simple form of the first non-zero term in 
Taylor expansion of action $S_\phi$. Let us mention that, in general, we 
will not rely on the expansion (\ref{TS}).  
There is another reason 
for choosing the combination $M^2M_{pl}$ in (\ref{action0}). Since the action 
is invariant under the ``shift'' symmetry $\phi\to\phi+\phi_0$, where $\phi_0$ 
is an arbitrary constant, the energy density is insensitive to the value 
of $\phi$. It is, however, sensitive to the values of derivatives of $\phi$.
Thus it is natural to make the $\Box$-operator dimensionless using one
scale, whereas the value of the field $\phi$ is made dimensionless 
with some other scale. The scale
$M$ can be viewed as the physical cutoff scale in our model. The other scale 
with which $\phi$ is made dimensionless we are free to choose at our 
convenience. We will choose this scale to be a Planck scale. With this choice,
it follows from the Friedmann equation  
that the Hubble parameter is naturally measured in units of $M$.    

Let us return to Eq.~(\ref{action0})
and perform the following substitution:
\begin{equation}
\phi\to M_{Pl}\phi,\ x^{\mu}\to M^{-1} x^{\mu}.
\label{newVariables}
\end{equation}
Note that the space-time coordinates and the Hubble parameter $H$ are made dimensionless with the same scale.
The operator $\Box$ is also 
dimensionless in the new variables due to the relation 
$\Box_{x}\to M^2 \Box_{x}$.
Thus the action (\ref{action0}) can be written in terms of the new dimensionless  
variables as:
\begin{equation}
  S \equiv S_g+S_{\phi}= 
  \left(\frac{M_{Pl}}{M}\right)^2\int d^4x\sqrt{-g}
  \left[-\frac{R}{2}+Q\left(\Box \phi\right)\right].
  \label{actionRescale1}
\end{equation}
We will further use a different action:
\begin{equation}
  S \equiv S_g+S_{\phi}= 
  \int d^4x\sqrt{-g}
  \left[-\frac{R}{2}+Q\left(\Box \phi\right)\right],
  \label{actionRescale}
\end{equation}
because the overall factor 
$(M_{Pl}/M)^2$ is irrelevant for the classical evolution.
This form of action is useful for further numerical analysis, since everything is 
dimensionless. To recover physical values of the Hubble parameter, energy density
etc. from our numerical results one has to perform a straightforward inverse procedure
\begin{equation}
\e\to M^2_{Pl}M^2\e,~~ H\to MH,~~ t\to M^{-1}t,... 
\end{equation}
where dots represent the rest of the physical parameters which are relevant for our model.
In this  paper we will work with the dimensionless variables defined in 
Eq.~(\ref{newVariables}) and action (\ref{actionRescale}) keeping the same notations for 
them as we had for dimensionfull variables.
This will be assumed in every expression unless specified otherwise.

\subsection{General formalism}
The equation of motion for the system described by the action 
(\ref{actionRescale}) can be obtained by the variation of $S$ over $\phi$:
\begin{equation}
   \frac{\delta S_{\phi}}{\delta \phi}\equiv
Q''\Box^2\phi+Q'''\left(\nabla_{\mu}\Box\phi\right)\left(\nabla^{\mu}\Box\phi\right)=0,
  \label{h1}
\end{equation}
where $(')$ means a derivative taken with respect to $\Box\phi$.
It is convenient to rewrite Eq.~(\ref{h1}) in the following form: 
\begin{eqnarray}
\Box B &=& -\left[\ln Q''(B)\right]'\left(\nabla_{\mu}B\right)\left(\nabla^{\mu}B\right)
,\nonumber\\
\Box\phi &=& B.
\label{h}
\end{eqnarray}
This form will be useful for the further analysis. The system (\ref{h}) is 
hyperbolic for an arbitrary function $Q(B)$, thus, the Cauchy problem is well posed. 
In particular, this means that the solutions 
obtained for the ideal homogeneous and isotropic Friedmann   universe  
are stable with respect to high frequency cosmological perturbations
\footnote{It is instructive to compare this situation with 
that for general "$k$-essence/ inflation'' 
models where requirements of stability provide restrictions on the possible Lagrangians 
(see \cite{k_inf,dil}).}.                                    

The energy-momentum tensor corresponding to the action (\ref{actionRescale}) is
\begin{eqnarray}
T_{\mu\nu}^{Q} &\equiv& \frac{2}{\sqrt{-g}}  \frac{\delta S_{\phi}}{\delta g^{\mu\nu}}=  
g_{\mu\nu}\left[Q'\Box\phi-Q+Q''
\left(\nabla_{\lambda}\phi\right)\left(\nabla^{\lambda}\Box\phi\right)
\right]-\nonumber\\
&-&Q''\left[\left(\nabla_{\mu}\phi\right)\left(\nabla_{\nu}
\Box\phi\right)+\left(\nabla_{\nu}\phi\right)\left(\nabla_{\mu}\Box
\phi\right)\right],
\label{met}
\end{eqnarray}
It is worthwhile noting that the Energy-Momentum Tensor (EMT) does not have the structure of
EMT of a perfect fluid. 

\subsection{Dynamics in the Friedmann Universe}
As usual in inflationary cosmology, let us consider the cosmological evolution 
of the system for  spatially flat Friedmann universe with the metric
\begin{equation}
ds^2=dt^2-a^2(t)d\mathbf{x}^2.
\label{metric}
\end{equation} 
The system of the equations of motion (\ref{h1}) can be written for this background 
in the following normal form
\begin{eqnarray}
\label{system_First}
\frac{d\dot B}{dt}&=&-3H \dot B-\left[\ln Q''(B)\right]'\dot B^2,\nonumber\\
\frac{dB}{dt}&=&\dot B,\\
\frac{d\dot \phi}{dt}&=&-3H \dot \phi+B,\nonumber
\end{eqnarray}
where $H=\dot a/a$. The first 
two equations integrate out to give
\begin{equation}
\label{converge}
Q''(B)\dot B={const\over a^3(t)}\rightarrow 0.
\end{equation}
In particular case of  $Q(B)=B^2/2$, the last 
expression yields that $B$ converges to some constant value $B_*$ 
which is not necessary equal to zero. It is precisely this 
peculiar fact that leads to inflation in the model under consideration.
In this case the energy-momentum 
tensor (\ref{met}) has a late time asymptotic of a  $\Lambda$ term.
Namely it is $T_{\mu \nu}=g_{\mu \nu}B_*^2/2$.  
One can expect that for a convex function $Q(B)$ with the 
curvature $Q''$ restricted from below the asymptotic
behavior is similar. Below we will see that this guess is correct 
and will discuss the underlying reason for that.  

\begin{figure}[t]
\begin{center}
\psfrag{B}[bl]{\small$B$}
\psfrag{fidot}[bl]{\small$\dot \phi$}
\psfrag{bdot}[bl]{\small$\dot B$}
\psfrag{fidot-}[t]{\small$\dot \phi_*<0$}
\psfrag{fidot+}[bl]{\small$\dot \phi_*>0$}
\includegraphics[width=4in]{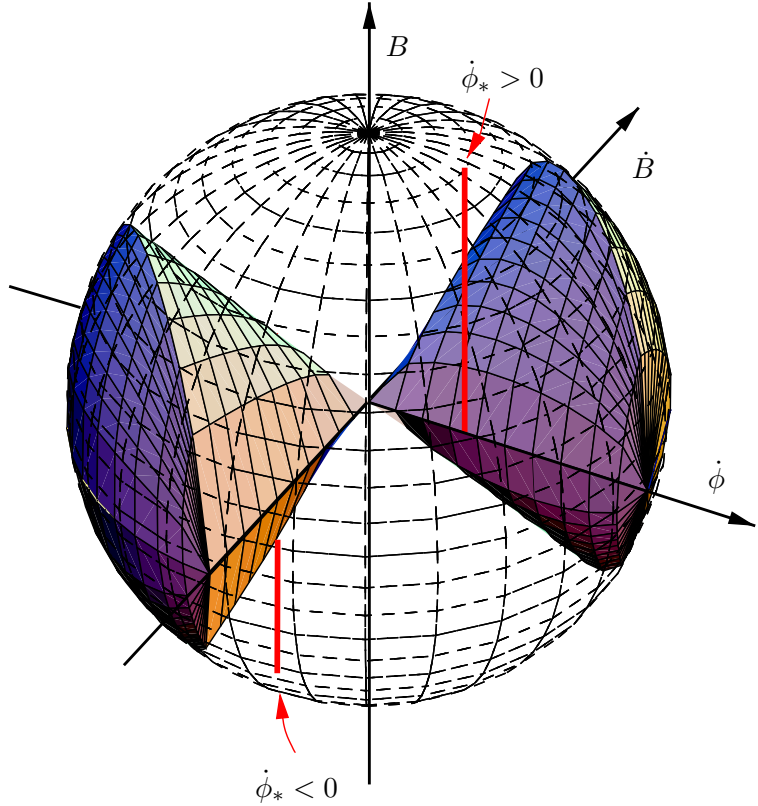}
\caption{\label{PhaseSpace}Here we plot a spherical region around the origin of the phase space
$(\dot\phi,\dot B,B)$. The colored volume represents the region where 
energy density $\e$ is negative.  
The surface of this volume 
which does not belong to the surface of the depicted sphere is the 
surface of zero energy density $\e=0$. Note that this forbidden regions 
are in the quarters of the $(\dot \phi, \dot B)$ 
plane where the $\dot \phi\dot B\geq0$
and consequently $w\leq-1$. The red curves represent the two fixed lines given by Eq.~(\ref{fixed}).}
\end{center}
\end{figure}

Up to now the system under consideration
is purely kinetic because neither equation of motion (\ref{h1}) 
nor the energy-momentum 
tensor (\ref{met}) explicitly  depend on $\phi$. 
Thus, in the case of the spatially flat Friedmann universe, 
the only relevant dynamical variables are $(\dot \phi, \dot B, B)$.  
In order to complete the description of the cosmological 
dynamics of this system we need
the Friedmann equation. As we have already mentioned the energy-momentum 
tensor (\ref{met}) does not have 
the algebraic structure of a perfect fluid one. 
Nevertheless in the homogeneous and isotropic case 
of the Friedmann universe the only non-vanishing components are 
diagonal and one can formally consider 
the energy-momentum tensor as if  it were of a perfect fluid. 
(Note, however,that the difference between these structures 
is important for the consideration of the more subtle questions like 
cosmological perturbations).  
Thus we can formally define the energy density $\e$ and pressure $p$ as:
\begin{eqnarray}
\label{EP}
\e&\equiv&T^0_{~0}=(Q'B-Q)-Q''\dot\phi\dot B,\\
p&\equiv&-\frac{1}{3} T^i_{~i}=-(Q'B-Q)-Q''\dot\phi\dot B.\nonumber
\end{eqnarray}    
Using these definitions we can write the Friedmann equation in the familiar form  
\begin{eqnarray}
H^{2} & = &\frac{\e}{3}  =
\frac{1}{3}\left[Q'(B)B-Q(B)-Q''(B)\dot{\phi}\dot{B}\right].
\label{Friedmann}
\end{eqnarray}
The equation of motion (\ref{system_First}) and the Friedmann equation (\ref{Friedmann}) yield 
the self consistent minimal set of equations which determines 
the cosmological evolution of the whole system -- scalar field $\phi$ + 
Gravity represented by $a(t)$. 
We will later refer to this case as the {\it self-consistent case}. 
The system is of the third order and the phase space is $(\dot \phi, \dot B, B)$.
As it follows from the Friedmann equation (\ref{Friedmann}) the energy density
$\e$ is restricted to be nonnegative. This provides the inequality  
$(Q'B-Q)\geq Q''\dot\phi\dot B$ which defines the region in the phase space
$(\dot\phi,\dot B, B)$ where the cosmological evolution is allowed 
(see Fig.~{\ref{PhaseSpace}}).

One can see from Eq.~(\ref{EP}) that the contributions to the energy density 
and the pressure
of a system can be divided into two parts: i) part $(Q'B-Q)$ which 
corresponds to the ``{\it cosmological-term}'' contribution since 
this term enters the expressions for $\varepsilon$ and $p$ with different signs; 
ii) part $-Q''\dot\phi\dot B$ corresponding to ``{\it ultra-hard}'' 
contribution. Note, that the ``cosmological-term'' part depends only on 
$B$ while ``ultra-hard'' part depends on $B$, $\dot\phi$ and $\dot B$. 
The presence of the cosmological-term-like part in the expressions for 
$\varepsilon$ and $p$ is a peculiar characteristic of the Lagrangian 
$Q(\Box \phi)$. In order to obtain a positive ``{\it cosmological-term}'' 
contribution to the energy density we should take only 
such functions into consideration
for which $(Q'B-Q)\geq 0$. The last inequality is always true for convex 
functions $Q(B)$. In this case the effective equation of state
$w\equiv p/\varepsilon$ is restricted to be less than one.
It is worthwhile noting that in the quarters of the phase space where  $\dot \phi\dot B\geq0$
the system is in phantom \cite{Phantom} regime: $w\leq-1$.  

\begin{figure}[t]
\begin{center}
\psfrag{t}[tr]{\small$t$}
\psfrag{f}[bl]{\small$\phi$}
\psfrag{fD}[bl]{\small$\dot\phi$}
\psfrag{b}[bl]{\small$B$}
\psfrag{bD}[bl]{\small$\dot B$}
\psfrag{fD*}[bl]{\small$\dot\phi_*$}
\psfrag{b*}[bl]{\small$B_*$}
\includegraphics[width=\textwidth]{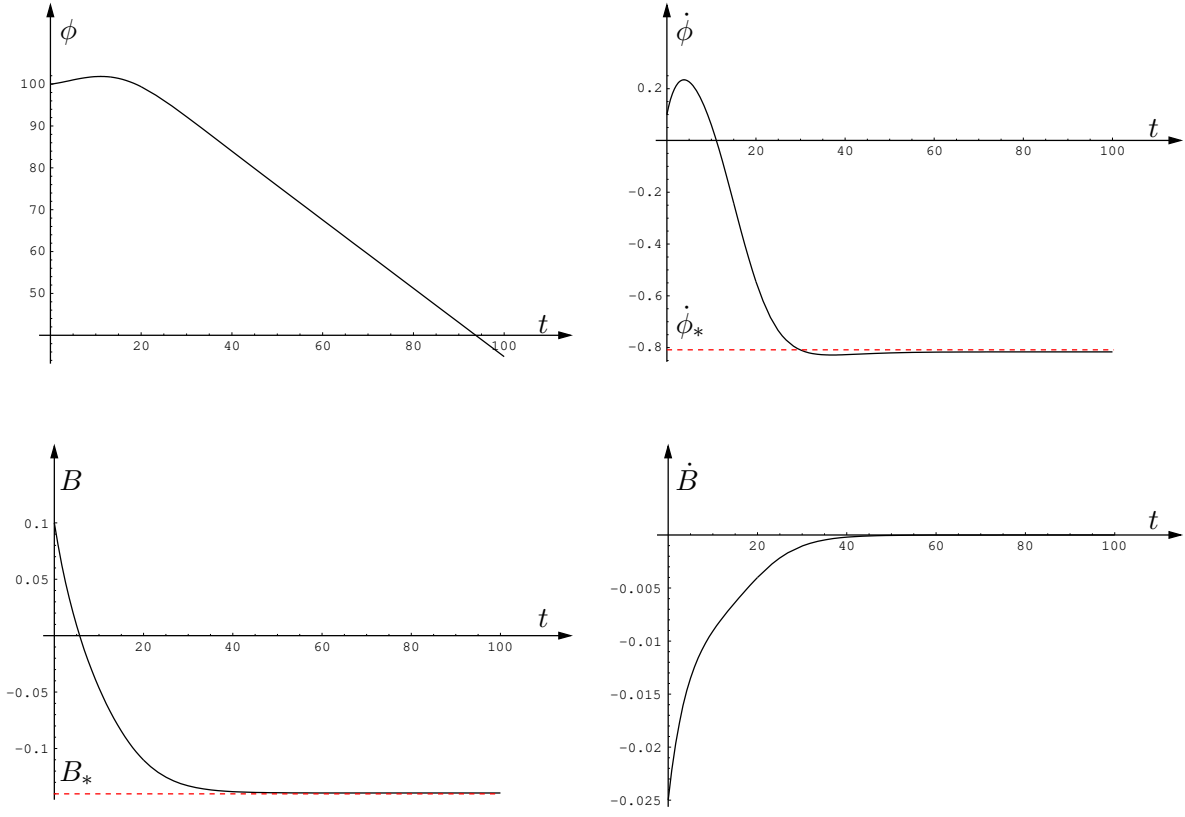}
\caption{\label{phiBInfl}
The evolution of the dimensionless parameters $\phi$ and 
$B$ is plotted versus dimensionless time $t$. Function $Q(B)$ is chosen to 
be $B^2/2$. The initial values are $\dot\phi=0.1$, $B=0.1$, $\dot B=-0.025$}
\end{center}
\end{figure}   

The system of equations (\ref{system_First}) where the Hubble parameter is 
expressed through the dynamical variables $(\dot\phi,\dot B, B)$ via the Friedmann
equation, possesses two kinds of fixed points. Namely a trivial one: 
$(\dot\phi,\dot B, B)=(0,0,0)$ which is of no interest for our purposes
and a nontrivial set of fixed points which form the two curves given by the equations
\begin{eqnarray}
  \label{fixed}
  \dot B=0,~B=B_*,~
  \dot \phi_*=\frac{B_*}{3H(B_*)}=\frac{B_*}{\sqrt{3(Q'(B_*)B_*-Q(B_*))}}.
\end{eqnarray}
Note that at each of these fixed points except $B_*=0$ the system 
is in the exact de Sitter inflationary regime. 
Further we will always assume that $B_*\neq0$.

\begin{figure}[t]
\begin{center}
\psfrag{N}[br]{\small$N$}
\psfrag{w}[bl]{\small$w$}
\includegraphics[width=4in]{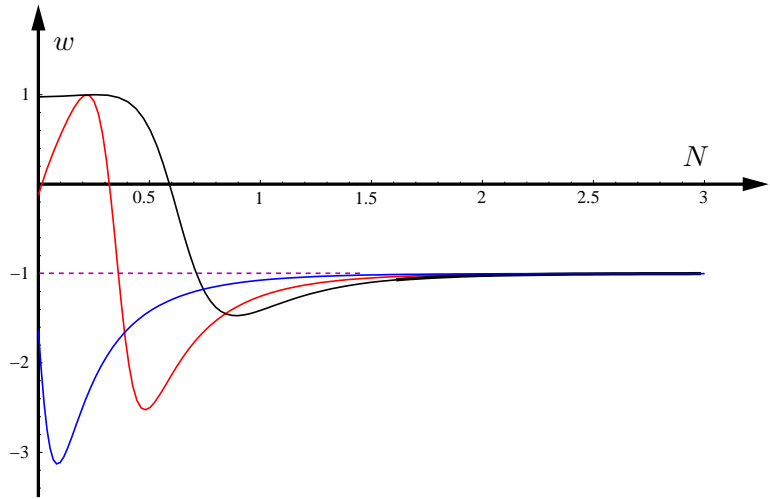}
\caption{\label{w(N)Enter}
Here we plot the equation of state $w$ versus number of e-folds 
$N$ for various 
initial data. The function $Q(B)$ is chosen to be $\frac{1}{2} B^2$.
The initial date are: for the red curve: 
$\dot\phi=0.1$, $B=0.08$, $\dot B=-0.05$, for the black curve:  
$\dot\phi=2$, $B=0.05$, $\dot B=-0.05$
for the blue one:  $\dot\phi=0.03$, $B=0.5$, $\dot B=1$. 
The equation of state $w$ evolves to the nearly de Sitter value $w=-1$ 
within approximately one e-fold.}
\end{center}
\end{figure}

Let us now investigate the character of these fixed points.
First we linearize the system (\ref{system_First}) around a given fixed point 
corresponding to a particular $B_*$. The linearized system is 
\begin{eqnarray}
  \label{linearized}
  \frac{d\dot B}{dt}&=&-3H_* \dot B \nonumber,\\
  \frac{d\delta B}{dt}&=&\dot B,\\
  \frac{d\delta \dot \phi}{dt}&=&-3H_*\delta \dot \phi + 
  \left[1-3H_* \gamma \right]\delta B +\gamma \dot B,  \nonumber
\end{eqnarray}
where we have defined
\begin{equation}
  \gamma=\frac{Q''(B_*) \dot \phi_*^2}{2H_*}, 
\end{equation} 
and used $\delta \dot B= \dot B$. One can find the eigenvalues 
corresponding to the matrix of the linearized system:
\begin{eqnarray}
  \label{eigenvalues}
  \lambda_1=0,~ \lambda_{2, 3}=-3H_*.
\end{eqnarray}
Because of the vanishing $\lambda_1$ the Lyapunov theorem is 
not applicable to the problem and we have to investigate the stability of the system 
in another way. Despite the failure of the linear analysis to prove the stability 
it is interesting to 
find a solution of the linearized problem.
The first two equations of the system (\ref{linearized}) do not depend on the 
$\delta \dot \phi$ and one can integrate them separately. The solutions are
\begin{eqnarray}
  \label{solutionsFirstTwo}
  \dot B =\dot B_0 \exp{(-3H_*t)},\nonumber \\
  \delta B =\delta B_0 +\frac{\dot B_0}{3H_*}\left[ 1-\exp{(-3H_*t)} \right],
\end{eqnarray}    
where $\dot B_0$, $\delta B_0=B_0-B_*$ are initial values for the linearized problem.
From these solutions it is clear that the stability in the usual sense cannot be established
because  the system is degenerate and the fixed points form a curve. 
Thus one can expect 
that after a small deviation from a given fixed point $B_*$ the system does not return 
to the same point $B_*$ but rather to a new point $B_{**}$. 
The natural question which arises is whether the new point $B_{**}$ 
is close to the original one.
In fact from Eq.~(\ref{solutionsFirstTwo}) we obtain the late time asymptotic 
\begin{equation}
  B_{**}= B_0 +\frac{\dot B_0}{3H_*}.
\end{equation}
\begin{figure}[t]
  \begin{center}
    \psfrag{t}[br]{\small$t$}
    \psfrag{p}[bl]{\small$p$}
    \psfrag{E}[bl]{$\varepsilon$}
    \psfrag{p*}[tl]{\small$p_*$}
    \psfrag{E*}[tl]{$\varepsilon_*$}
    \psfrag{w}[bl]{\small$w$}
    \includegraphics[width=\textwidth]{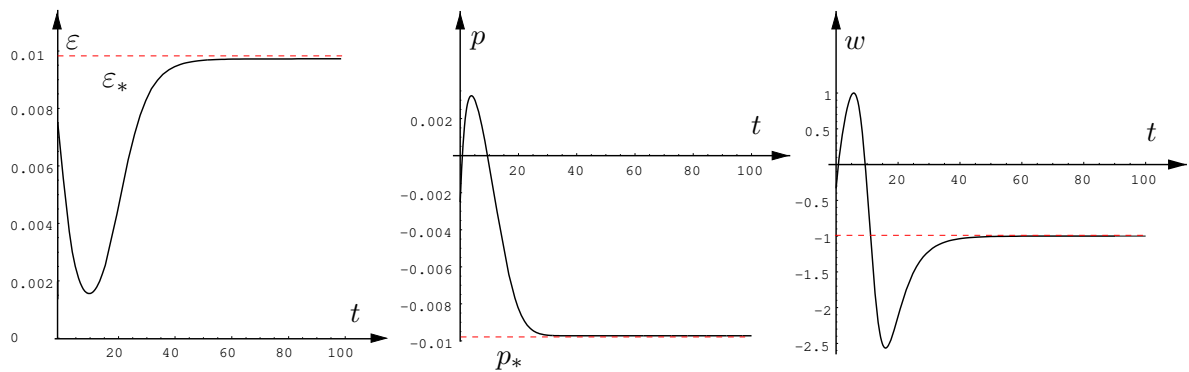}
    \caption{\label{EPInfl}
      The evolution of the energy density $\varepsilon$, 
      pressure $p$ and the equation of state $w=p/\varepsilon$ is shown on this 
      plot.  Function $Q(B)$ is chosen to be $B^2/2$. The initial values are 
      $\dot\phi=0.1$, $B=0.1$, $\dot B=-0.025$. The system quickly evolves into 
      a nearly de Sitter regime.}
  \end{center}
\end{figure}  
Therefore $\delta B_*=B_{**}-B_{*}=\delta B_0 + \dot B_0 / 3H_*$ and for $\dot B_0/3H_*\ll1$ 
this deviation is approximately equal to the initial deviation: $\delta B_*\simeq \delta B_0 $.
However, as one can directly prove the linearized solution for $\delta \dot \phi$
does not converge to the fixed point $\delta \dot \phi=0$,
if the point of linearization $B_*$ does not coincide with late 
time asymptotic $B_{**}$. 
But if we linearize the system around the value of the late time asymptotic 
$B_0 + \dot B_0 / 3H_*$, 
then the solution for $\delta \dot \phi$ is
\begin{equation}
  \delta\dot \phi=\left[\delta \dot \phi_0+t \dot B_0 (2\gamma-3 H_*) \right]\exp{(-3H_*t)}.
\end{equation} 
Thus we prove that in the linear approximation the solutions converge to the curves formed by 
the fixed points (\ref{fixed}). The validity of the linearized solutions can be numerically  
confirmed for the appropriately chosen initial data.  
The numerical calculation  reveals the following picture. The 
generic evolution for $\phi$, $\dot\phi$, $B$ and $\dot B$ 
is shown in Fig.~\ref{phiBInfl}. 
One can see from Fig.~\ref{EPInfl} that, indeed, the asymptotic solution is 
inflationary. The initial conditions which lead to this sort of behavior are 
rather general. The system reaches the inflationary regime relatively quickly  
(see Fig.~\ref{w(N)Enter}) and will stay there forever
if nothing else is added to the action (\ref{action0}). 
Of course, in the real world there has to be a mechanism to exit from this 
inflationary regime. This will modify the evolution shown in 
Figs.~\ref{phiBInfl} and~\ref{EPInfl}.
We will propose a natural modification and discuss the details  of the 
graceful exit in  the next section,
while below we will complete the study of the system described by 
(\ref{action0}).   

\subsection{The Behavior of $\left(\varepsilon,p\right)$ Trajectories}
In order to develop intuition it is 
useful to consider the trajectories of the system in 
$\left(\varepsilon,p\right)$ plot. 
A state of the system is not uniquely described by a point 
in the $\left(\varepsilon,p\right)$ plot.
As we have already mentioned, to solve the full system of the 
equations (Friedmann equation (\ref{Friedmann}) 
and equation of motion (\ref{system_First}) one has to specify the 
following initial conditions: 
$\dot\phi_0$, $B_0$ and $\dot B_0$. In fact, we can replace these 
initial conditions by specifying the initial energy density $\e_0$, 
the initial pressure $p_0$ and the initial angle of the  
$\left(\varepsilon,p\right)$ -- trajectory:
$\tan\theta=dp/d\varepsilon=\dot p/\dot\varepsilon$. 
It is worthwhile noting that if our system 
had a perfect fluid like EMT, then the introduced angle $\theta$ 
would be closely connected to the adiabatic speed of sound: $\tan\theta=c_s^2$. 
From the stability requirements it would  follow $\tan\theta\geq0$. Our system is free 
from this restriction due to the hyperbolicity of the system (\ref{h}).  
Using the equation of motion (\ref{system_First}) we can rewrite the formula for the 
initial angle in the following form
\begin{equation}
\label{tan}
\tan\theta=1-\frac{B}{3H\dot\phi}\ .
\end{equation}
Note that the Eq.~(\ref{tan}) along with the continuity equation
\begin{equation}
\label{contin}
\dot\e=-3 H (\e+p),
\end{equation}
uniquely define the angle $\theta$.
One can easily check that $(\dot\phi,\dot B,\ B)\to(\e,\ p,\ \theta)$
is one-to-one correspondence for all finite initial values except when
 $\dot\phi_0=0$.

\begin{figure}[t]
\begin{center}
\psfrag{E}[br]{$\varepsilon$}
\psfrag{p}[bl]{\small$p$}
\psfrag{Te}[bl]{\small$\theta_{cr}$}
\psfrag{p=E}[br]{\small$p=\varepsilon$}
\psfrag{p=-E}{\small$p=-\varepsilon$}
\psfrag{p=-1/3E}{\small$p=-\frac{1}{3}\varepsilon$}
\psfrag{0}[br]{$0$}
\includegraphics[width=4in]{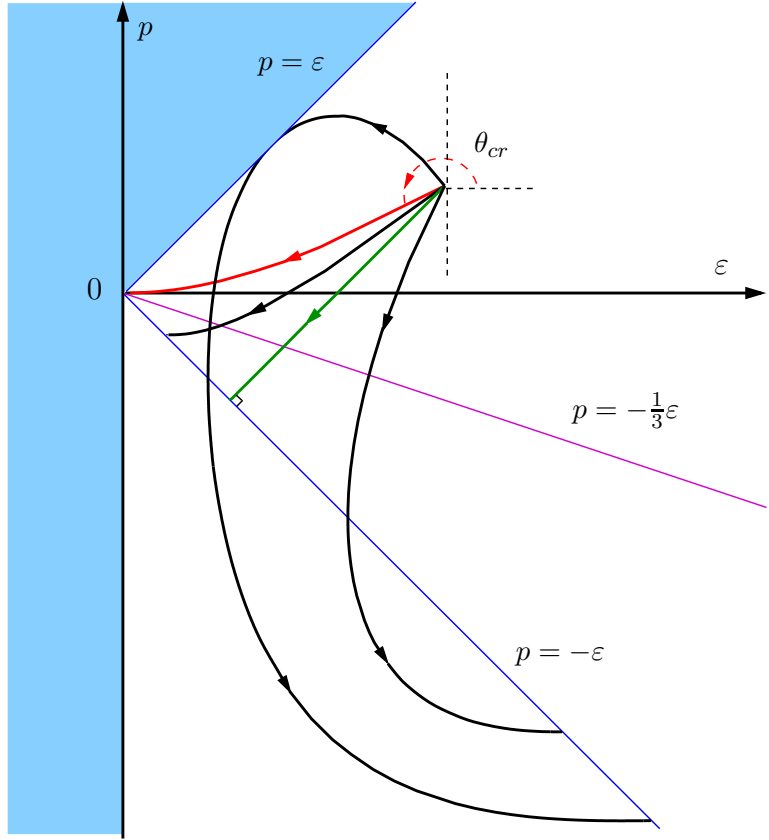}
\caption{\label{phase1} Here we plot the $(\e,p)$-trajectories originating 
from a point where the universe is decelerating. 
The blue region cannot be reached by 
the system.}
\end{center}
\end{figure}
Figure \ref{phase1} represents
trajectories starting with different initial angles 
$\theta_0$ at the same point $(\e_0,\ p_0)$ which is located in the 
region ($p_0>-\e_0/3$), where the universe is decelerating.
Let us analyze the behavior of the trajectories. 
First we notice that in virtue of the continuity equation (\ref{contin}) 
a trajectory can be vertical only at the point of  
the crossing of the ''phantom divide" $p=-\e$ 
\footnote{The possibility of a transition through the phantom divide  
was discussed e.g. in Refs. \cite{alex,Quintom,Zhang}}.
Namely, by the crossing from the non-phantom side we have $\theta=3\pi/2$, and by the crossing 
from the phantom side $\theta=\pi/2$.   
If the system has a de Sitter solution as a late time asymptotic, 
then $\theta=0$ by approaching from the phantom side and $\theta=\pi$ otherwise, as it follows 
from Eq. (\ref{fixed}).
On the other hand from  Eq.~(\ref{tan}) we find  that $\tan\theta=1$ 
can be achieved with finite values of the dynamical variables only if
$B=0$. While from the definition (\ref{EP}) of the energy density we have $p= \e$ in this case.
Thus the line $p=\e$ corresponding to the ultra-hard equation of state is the only 
possible place in the $(\e,\ p)$ plot where $\tan\theta$ can be equal to one along with 
the values  $(\dot\phi,B,\dot B)$ being finite. 
Therefore all trajectories reaching the ultra-hard line $p=\e$ at $\e \neq 0$ 
have the ultra-hard line as a tangent line at this point.  
It is clear that the trajectories have the angle $5/4 \pi$ when 
touching the ultra-hard line. 
If a trajectory touches this line away from the origin $(\e,\ p)=(0,0)$ 
then $\dot B\dot \phi\neq0$ 
and  the only possibility for a further evolution is to reflect in the direction of the 
smaller $\e$. Immediately after the reflection the trajectory can only increase its angle. 
 The angle  is now bounded from below with $\theta=5/4 \pi$ and bounded 
from above by $\theta=3/2 \pi$. Thus, moving downwards the trajectory increases its 
angle and inevitably crosses the phantom divide.  
If we take into account that the angle $\theta$ must be a continuous function, we find that all
trajectories which cross the phantom divide and which have the initial angle $\theta< \pi$ 
must reflect from the ultra-hard line.
Another possibility to have $\tan \theta=1$ corresponds to infinite 
values of $\dot B$ and $\dot \phi$. 
This is a limiting case which is shown by the green line in  Fig.~(\ref{phase1}). 
For the trajectories with the initial angles 
large than $\theta=5/4 \pi$ the angles are bounded from below and from above  
and the trajectories are evolving downwards. 
Thus, they must cross the phantom divide in any case. 
Considering a limit of the trajectories which touch the ultra-hard line 
we find that there must exist a critical trajectory which goes to the $(\e,\ p)=(0,0)$.
For a further consideration of  it is convenient to introduce 
the critical angle $\theta_{cr}= \theta_{cr}(\e_0,\ p_0)$ which
corresponds to the critical trajectory having the 
late time asymptotic:  $\e\to 0$, $p\to 0$ as $t\to\infty$.
This critical trajectory is shown in red in Fig.~{\ref{phase1}}.
Then  all the trajectories can be divided into three subclasses. 

The first subclass contains trajectories with the initial angles
$\pi/2<\theta<\theta_{cr}$. The properties of these trajectories are the 
following: they go upwards, (if $p_0<0$ intersect the axis $p=0$ at some point), then reflect 
from the line $p=\e$ at the angle $5\pi/4$. After that they again
intersect the axis $p=0$ and go to the phantom regime crossing $p=-\e$ at 
the angle $3\pi/2$. Finally, these trajectories  approach the de Sitter
regime $p=-\e$.

\begin{figure}[t]
\begin{center}
\psfrag{E}[br]{$\varepsilon$}
\psfrag{p}[bl]{\small$p$}
\psfrag{Te}[bl]{\small$\theta_{cr}$}
\psfrag{p=E}[br]{\small$p=\varepsilon$}
\psfrag{p=-E}{\small$p=-\varepsilon$}
\psfrag{p=-1/3E}{\small$p=-\frac{1}{3}\varepsilon$}
\psfrag{0}[br]{$0$}
\includegraphics[width=4in]{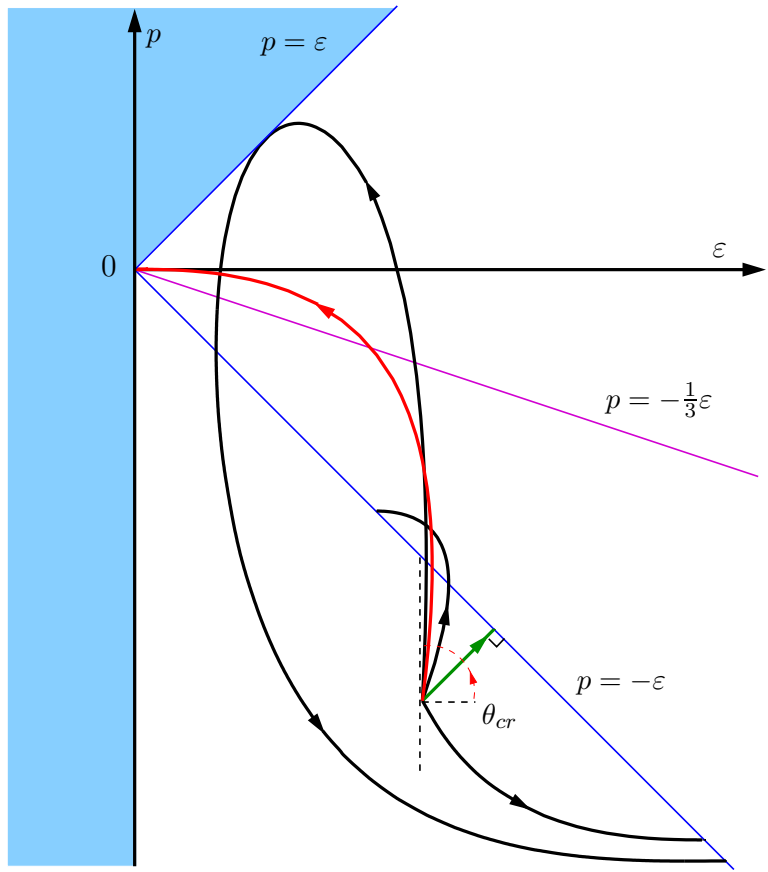}
\caption{\label{phase2} Here we plot the $(\e,p)$-trajectories originating 
from a point where the universe undergoes a super-acceleration stage. 
The blue region cannot be reached by 
the system.}
\end{center}
\end{figure}

The second subclass contains trajectories with the initial angle 
$\theta_{cr}<\theta<5\pi/4$. In this case trajectories go
directly to $p=-\e$, approaching this line as $t\to\infty$ but never crossing it.
This happens because these trajectories are moving downward towards the phantom divide
with the angle bounded from above by $5/4\pi$. A trajectory cannot have this limiting 
angle because it evolves away from the ultra-hard line and in that case the $5/4\pi$ angle 
corresponds to an infinite $\dot \phi$ or $\dot B$.    

And, finally, the third subclass contains the trajectories with the initial 
angle $5\pi/4<\theta<3\pi/2$. These curves intersect $p=-\e$ at 
the angle $3\pi/2$ going to the phantom regime and then approach the 
inflationary regime $p=-\e$. 

Figure \ref{phase2} shows
trajectories which begin in the phantom regime ($\varepsilon_0
+p_0<0$) with different initial angles $\theta_0$. 
In this case the analysis is analogous to 
the previous case. The whole set of 
trajectories 
also can be divided into three subclasses as in the case of the ``ordinary''
initial conditions ($\ p_0>-\e_0/3$). The critical angle $\theta_{cr}$ is 
also defined as the initial angle of the trajectory which 
approaches $\e=0$, $p=0$ as $t\to\infty$. The critical 
trajectory is shown in red in the Fig.~\ref{phase2}.
The first subclass consists of trajectories 
with initial angles $\theta_{cr}<\theta<\pi/2$. In this
case trajectories intersect $p=-\e$ at the angle $\pi/2$. Then
one arrives to the situation which was described by the first subclass 
in $(\e_0+p_0>0)$ case. The second subclass is the trajectories, which 
start with the angles $\pi/4<\theta<\theta_{cr}$. The trajectories intersect
the line $p=-\e$ and then one starts from the conditions described 
in the second subclass of  $(\e_0+p_0>0)$ case. The third subclass is the
set of trajectories with the initial angles $-\pi/2<\theta<\pi/4$. These are 
the curves which directly start to approach the inflationary regime 
$p=-\e$. 

By the analysis performed in this subsection we confirm our result 
that the system generically approaches the de Sitter regime $p_*=-\e_*$. 
The only exceptions are the critical trajectories (shown in red), 
in this case $\varepsilon_*=0$. However as we have shown for each point $(\e_0, p_0)$ there 
is only one solution corresponding to the critical curve. Thus the solutions which do not 
lead to the late time de Sitter regime have a zero measure. 

\subsection{Stability of the system}

It is well known (see, for example, recent papers \cite{Higher}) that higher derivative
theories possess ghost (or phantom). The presence of a ghost degree of freedom 
in a dynamical system leads usually to various concerns about classical 
and quantum stability of such a system \cite{Cline}. While the detailed investigation
of the relevancy of all these problems regarding our model is under way \cite{pert}
we would like, perhaps, to make a few comments on these issues below.

First of all, let us briefly discuss the classical stability of the model.
In the above analysis we have studied the dynamics governed by the homogeneous 
equations of motion only.
It was shown that the solutions of these equations are stable with respect to the perturbations 
in the initial conditions. However, 
showing that solution of the homogeneous equations is stable 
is not enough for cosmological applications. Even disregarding the perturbations of the metric, one has
to worry about the classical stability of this solution 
of the partial differential equations   
in $3+1$ space-time. Regarding this issue we can note that the equations in our model 
are hyperbolic. Therefore the Cauchy problem is well posed. 
This means that the dependence of the solutions on the initial conditions is continuous and 
exponential instabilities are excluded.  
Thus, possible classical instabilities, if any, are not as dramatic
as it is for systems which may flip from hyperbolic type to elliptic type in the course
of the dynamical evolution. 
If one includes the perturbations of the metric, the problem becomes even 
more involved and the further
analysis is needed.

Finally, one may also worry about quantum instability 
due to the one graviton exchange between the ghost degree of freedom and the Standard
Model particles. 
While during inflation the argument from  \cite{Cline} is not applicable, after the exit 
the residual ghost may pose such problem. There is no doubt, that weakly 
coupled ghost degree of freedom leads to the vacuum instability. However, if 
the ghost couples strongly to a regular scalar field in a particular way, which is the case in 
our model, the argument based on the evaluation of the creation rate of Standard Model particles 
out of vacuum is not reliable anymore. This issue certainly deserves further investigation. 



\section{Exit Mechanism}

Every realistic inflationary scenario needs to have a natural 
exit mechanism. In this part we will modify our model in order to 
have an exit from the inflationary regime. Let us return to the 
action (\ref{action0}) and add a first--order derivative term. It can be, for example,
the first term from the Lagrangian (\ref{mukh}).   
Thus we shall consider the following action for the
scalar field:
\begin{equation}
  S=\int d^4x\sqrt{-g}\left[ K\left(\frac{\phi}{M_{Pl}}\right) \frac{(\nabla_\mu\phi)^2}{2}+
  M^2 M_{Pl}^2\, Q\left(\frac{\Box \phi}{M^2 M_{Pl}}\right)\right].
  \label{action00}
\end{equation} 
This  choice has been widely discussed as an effective approach 
in string theory \cite{string} and its simplicity facilitates further analysis. 
In a cosmological context
such a term was considered in \cite{k_inf,dil}.

To simplify the analysis of the system we again use the dimensionless
variables defined according to (\ref{newVariables}).
The overall factor $(M_{Pl}/M)^2$  is irrelevant for the classical analysis 
and we can use the following action instead:
\begin{equation}
  S=\int d^4x\sqrt{-g}\left[K(\phi)X+
  Q(\Box \phi)\right],
  \label{action}
\end{equation}
where $X$ depends on the dimensionless 
quantities $x^\mu$ and $\phi$ as usual
$X=\frac{1}{2}(\nabla_\mu\phi)^2$.

The implementation of the exit from the inflationary regime 
is based on the following idea. 
We assume that there are two regions of $\phi$:
one is a $Q$-dominated region, where $\left|K(\phi)\right|\ll 1$,
so that the first term in (\ref{action}) does not 
influence the dynamics; and another is the region
where the term $K(\phi) X$ influences the dynamics significantly
in such a way that the system exits from the inflationary 
regime. We will call this region the $K+Q$-region.
Starting from $Q$-region, 
the system will enter 
the inflationary regime in a way very similar to 
that discussed in the previous section.
Then during inflation the value of $\phi$ is 
changing as $\phi(t)=\phi_0+\dot \phi_*t$ (where $|\dot\phi_*|\neq 0$ 
according to Eq.~(\ref{fixed})), 
and the system may come eventually to the $K+Q$-region.
For example, if $K(\phi)\to 0$ for $\phi\to -\infty$ and 
$K(\phi)\to \infty$ for $\phi\to +\infty$, then starting
from inflation in the region of large negative $\phi$
with $\dot\phi_*>0$, our system will eventually come to 
the $K+Q$-region, where we expect the system to exit from the 
inflation. During the evolution, the contribution to the 
action which comes from the $K$-term
grows during the inflationary regime
whereas the contribution from the $Q$-term remains almost the same.
Therefore we can roughly estimate the time when inflation ends:
$\dot \phi_*^2|K(\phi_e)|\simeq Q(B_*)$ where $\phi_e=\phi_0+\dot \phi_*t_e$.
The number of e-foldings during the inflationary regime is 
given by 
\begin{equation}
N\simeq H_*{|\phi_0-\phi_e|\over |\dot\phi_*|}.
\end{equation}
Below we will find the necessary conditions on the function $K(\phi)$ in order  
to have an exit from the inflationary regime,
and will show that a large class of functions $K(\phi)$ has the desired properties.

The energy density and the pressure 
coming from the additional term in the action $K(\phi)X$
give the following contribution to the energy density 
and the pressure of $Q$-part of Lagrangian (Eq.~(\ref{EP})):
\begin{eqnarray}
  \label{EP1}
  \e_1=p_1= K(\phi)X,
\end{eqnarray}  
where subscript $''1''$ reflects the fact that the term under consideration contains no 
higher order derivatives. 
The equations of motion (\ref{h}) become
\begin{eqnarray}
\Box B&=&-\left[\ln Q''(B)\right]'\left(\nabla_{\mu}B\right)^2+(Q''(B))^{-1}\left[K(\phi)B
+\frac{1}{2} K'(\phi)\left(\nabla_{\mu}\phi\right)^2\right]
,\nonumber\\
\Box\phi&=&B.
\label{eom}
\end{eqnarray}
For simplicity we assume $Q(B)=B^2/2$ in the further analysis, since 
there is no qualitative difference between this choice and other
convex functions $Q(B)$. 
The equation of motion (\ref{eom}) simplifies to
\begin{eqnarray}
\Box B&=&K(\phi)B+\frac{1}{2} K'(\phi)\left(\nabla_{\mu}\phi\right)^2
,\nonumber\\
\Box\phi&=&B.
\label{eomB^2}
\end{eqnarray}
These equations of motion have the structure similar to the usual 
Klein-Gordon equation. In particular the first term on the right hand side 
plays the role of a mass term for the field $B$. 
Using this analogy we come to the conclusion that in order to avoid a 
tachyonic instability we have to assume that $K(\phi)<0$. 
For the Friedmann universe the system of equations (\ref{eomB^2}) 
takes the following form 
\begin{eqnarray}
  \label{B_homog}
  \ddot B +3H\dot B=-m^2(\phi)B-m(\phi)m'(\phi)\dot \phi^2,\\
  \ddot\phi+3H\dot \phi=B,
  \label{phi_homog}
\end{eqnarray}
where we have introduced the effective mass $m^2(\phi)\equiv -K(\phi)$.
When the r.h.s.~of Eq.~(\ref{B_homog}) becomes large enough 
then inflation ends. The value of $\phi$ where the evolution 
changes significantly can be estimated by a comparison of the friction 
term in (\ref{B_homog}) and the ``mass term'', $m^2(\phi)B$. When 
\begin{equation}
  \label{Hsimm}
  H\sim m(\phi)
\end{equation} 
the field $B$ becomes massive and starts to oscillate and, thus, inflation ends.
This is in fact quite similar to the exit from inflation 
in ``standard'' {\it potential induced} scenarios. 

While the
behavior of the system in the transition regime between the box-dominating phase
and the oscillating phase  is difficult to study analytically, it is 
possible to describe the system in the asymptotic oscillating regime.
Our further analysis is based on the book \cite{Mukhanov} chapter 4.
Introducing the rescaled fields $u\equiv aB$, $v=a \phi$ and the conformal time 
$\eta=\int dt/a$, we can rewrite the system (\ref{B_homog}),(\ref{phi_homog}) in the 
following form:
\begin{eqnarray}
u''+\left(m^2a^2-\frac{a''}{a}\right)u&=&-\frac{v^2}{a}m\frac{dm}{d\phi}
,\nonumber\\
v''-\frac{a''}{a}v=u,
\label{EofMconf}
\end{eqnarray}  
where the prime denotes the derivative with respect to $\eta$.
The term on the right hand side in the first of these equations rapidly decays in the 
expanding Friedmann universe. Following the book \cite{Mukhanov} we find that if 
$\left|a''/a^3 \right|\sim H^2 \gg m^2$
the first term in the brackets can be neglected as well and the approximate solution corresponds 
to the frozen field $B=B_*$. This is exactly the inflationary solution obtained 
in the previous section.
In our problem the effective mass $m(\phi)$ is growing with time, whereas 
the Hubble constant changes only slightly. As the mass becomes larger than the Hubble 
parameter $H \sim H_*$ we can neglect the second term inside the brackets in 
Eq.~(\ref{EofMconf}).
The WKB solution of the simplified equation written in terms of $B$
is then 
\begin{equation}
\label{wkb}
B\propto a^{-3/2} \frac{1}{\sqrt{m(\phi)}}\sin \left[ \int m(\phi)dt \right].
\end{equation}
This solution is valid in the case of a slowly varying mass $m(\phi(t))$, i.e. if
\begin{equation}
{(ma)'_{\eta}\over m^2a^2}\ll 1.
\end{equation}
Going back from the conformal time units to a cosmic time the above inequality becomes
\begin{equation}
{\dot m\over m^2}+{H\over m}\ll 1.
\label{ineq}
\end{equation}
Since $H\ll m$ already, the first term in (\ref{ineq}) is also much less then one. 
From the equation (\ref{B_homog}) it follows that
\begin{equation}
\dot\phi={1\over a^3}\int dt Ba^3 \propto 
-a^{-3/2}m^{-3/2}(\phi)\cos\left[ \int m(\phi)dt \right]. 
\end{equation} 
Estimating the energy density we obtain
\begin{equation}
\e\propto {B^2\over 2}-\dot B\dot\phi-{m^2\dot\phi^2\over 2}\propto {1\over a^3}.
\label{energy} 
\end{equation}
The neglected next-to-leading terms we have neglected are of the order of 
$H(t)a^{-3}(t)$ and are oscillating. Since 
$H(t)$ is a decreasing function of time and we average over 
fast oscillations these terms, 
indeed, 
quickly become subdominant. 
One can also estimate that the pressure behaves as
\begin{equation}
p\propto -a^{-3}\cos\left[ 2\int m(\phi)dt \right]. 
\label{pressure}
\end{equation}
After averaging over fast  oscillations one obtains that $\langle p \rangle =0$. 
Thus our system has a dust-like behavior after the inflation is over.

Now we are ready to formulate the conditions for $K(\phi)$ in order to obtain 
the exit from inflation using the action (\ref{action}):i) $K(\phi)<0$; ii) there should 
exist the region of $\phi$ where $\left|K(\phi)\right| \ll 1$; iii) Eq.~(\ref{Hsimm}) 
must have a solution at some $\phi_e$,
where the value of $H$ is taken as it were in the inflationary regime.
The last condition means, in particular, that for a function $K(\phi)$ bounded
below, the presence of the exit in the model depends on the value of 
$B$ (and consequently on the value of $H_{inf}$) during the inflation. 
Below we study numerically two examples of the function 
$K(\phi)$ to illustrate some generalities of the proposed exit mechanism
\footnote{It is worthwhile noting that the discussed exit mechanism reveals 
some similarities with Ref. \cite{Piao}}.  

The exit from the inflation which we have constructed  
still has to be complemented by a reheating mechanism, which would produce 
radiation after the end of inflation. Since the residue component has 
a dust-like behavior after inflation ends, 
the mechanism which converts its energy into radiation is the same 
as in the standard inflationary scenarios: 
we need to couple the oscillating field (in our case it can be $B$ instead of
the inflaton field) to some scalar fields 
$\chi$ through, for instance, an interaction term $B\chi^2/\Lambda$ where $B$ and
$\chi$ have dimension three and one respectively, and 
$\Lambda$ can be a Planck or a somewhat lower scale.
Such an interacting term would lead to an energy density transfer 
from  the massive ``field'' $B$ to the radiation-like particles {\it via}, 
for example, the narrow (or broad) parametric resonance. We will discuss the details
in the forthcoming paper \cite{pert}. 

\begin{figure}[t]
\begin{center}
\psfrag{t}[br]{\small$t$}
\psfrag{f}[bl]{\small$\phi$}
\psfrag{fD}[bl]{\small$\dot\phi$}
\psfrag{b}[bl]{\small$B$}
\psfrag{bS}[bl]{\small$\dot B$}
\psfrag{fie}[br]{\small$\phi_e$}
\psfrag{B*}[br]{\small$B_*$}
\psfrag{fD*}[br]{\small$\dot\phi_*$}
\includegraphics[width=0.9\textwidth]{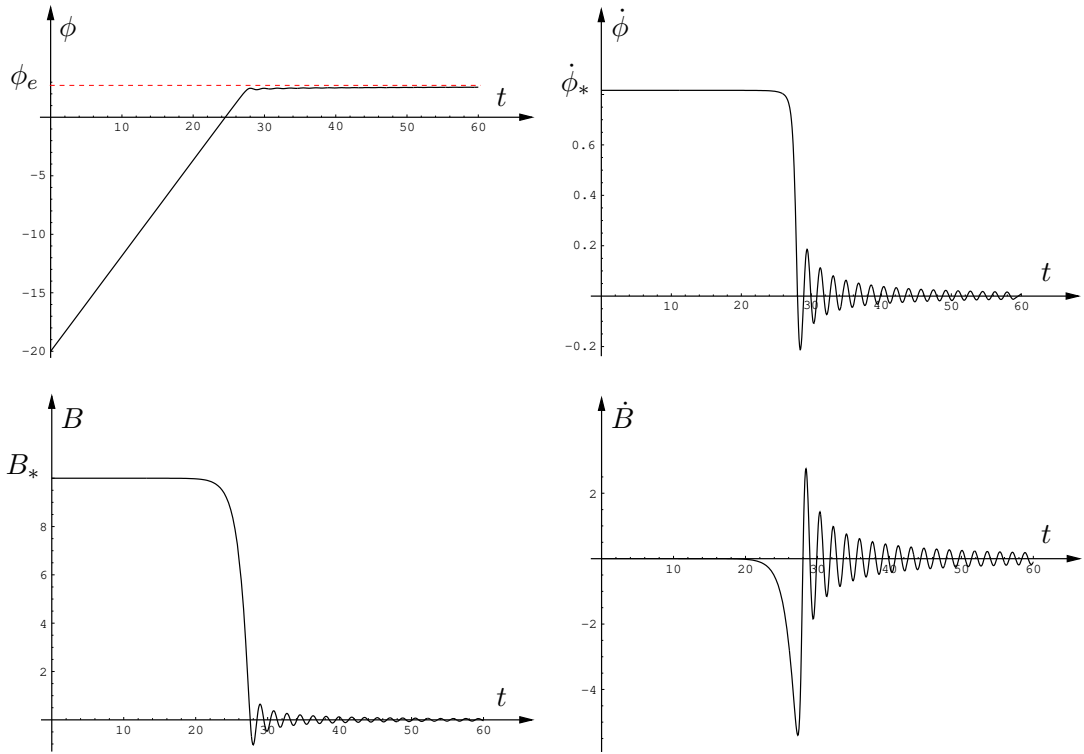}
\caption{\label{ex2a} Dynamics for the system 
$K(\phi)=-\exp \phi$, $Q=B^2/2$.
Initial values were $\phi=-20$, $\dot\phi=\dot \phi_*=\sqrt{2/3}$, 
$B=10$, $\dot B=0$.}
\end{center}
\end{figure}   

\begin{figure}[t]
\begin{center}
\psfrag{t}[br]{\small$t$}
\psfrag{p}[bl]{$p$}
\psfrag{E}[bl]{$\varepsilon$}
\psfrag{w}[bl]{\small$w$}
\includegraphics[width=3in]{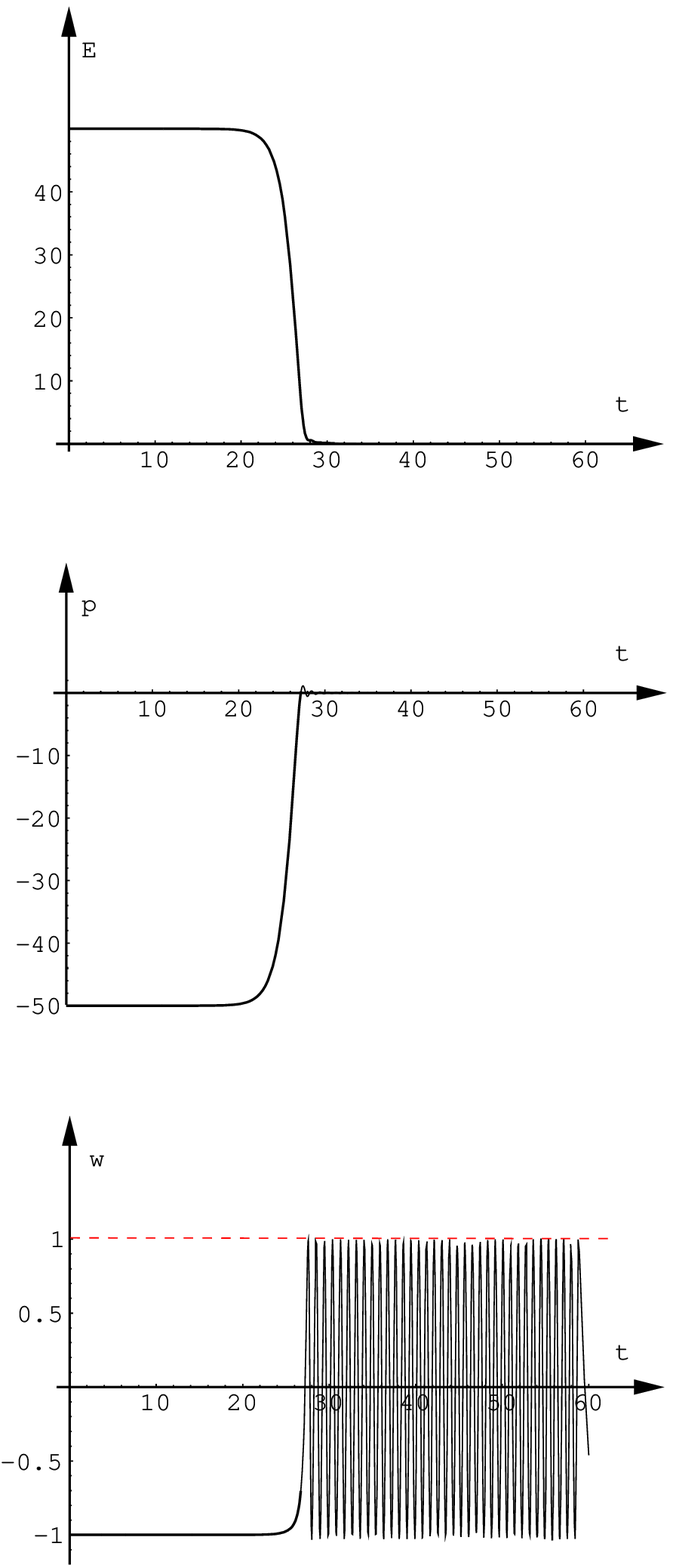}
\caption{\label{ex2b} Energy density, pressure and equation of state 
for the system $K(\phi)=-\exp \phi$, $Q=B^2/2$.
Initial values were $\phi=-20$, $\dot\phi=\dot \phi_*=\sqrt{2/3}$, 
$B=10$, $\dot B=0$.}
\end{center}
\end{figure}
\subsection{Numerical examples}

First we consider the case 
\begin{equation}
  \label{K1}
  K(\phi)=-e^{\phi}.
\end{equation}
We shall see below that the exit from the inflationary regime in this case 
is sharp in comparison with the next example. This may be important
for generation of the curvature fluctuations. This will, however, be discussed
in \cite{pert}.

For the choice (\ref{K1}) the inflationary regime starts for
large negative $\phi$. There are two possible sets of initial 
conditions: $\dot\phi_*<0$ and $\dot\phi_*>0$. 
If $\dot \phi_*>0$ then $\phi$ grows as $\phi=\phi_0+\dot\phi_*t$ in the inflationary 
regime and the term 
$-X\, e^{\phi}$ becomes more and more important and eventually 
drives the system into a state where the parameter $w$ 
oscillates as shown in Figs.~\ref{ex2a} and \ref{ex2b}. 
The time average of these oscillations is $w \simeq 0$, 
i.e. the system after the exit has a dust-like
equation of state. The behavior of the system for $t\to\infty$
can be described by Eqs.~(\ref{B_homog}) and (\ref{phi_homog}) 
with $m\simeq e^{\phi_e/2}$, where
$\phi_e$ can be roughly estimated using Eq.~(\ref{Hsimm}).
The energy and pressure are given by Eqs. (\ref{energy}) and (\ref{pressure}). 
In the case of $\dot \phi_*<0$ there is no exit from the inflationary 
regime for the chosen function $K(\phi)$.  

\begin{figure}[t]
\begin{center}
\psfrag{t}[br]{\small$t$}
\psfrag{f}[bl]{\small$\phi$}
\psfrag{fD}[bl]{\small$\dot\phi$}
\psfrag{b}[bl]{\small$B$}
\psfrag{bS}[bl]{\small$\dot B$}
\psfrag{B*}[br]{\small$B_*$}
\psfrag{fD*}[br]{\small$\dot\phi_*$}
\includegraphics[width=0.9\textwidth]{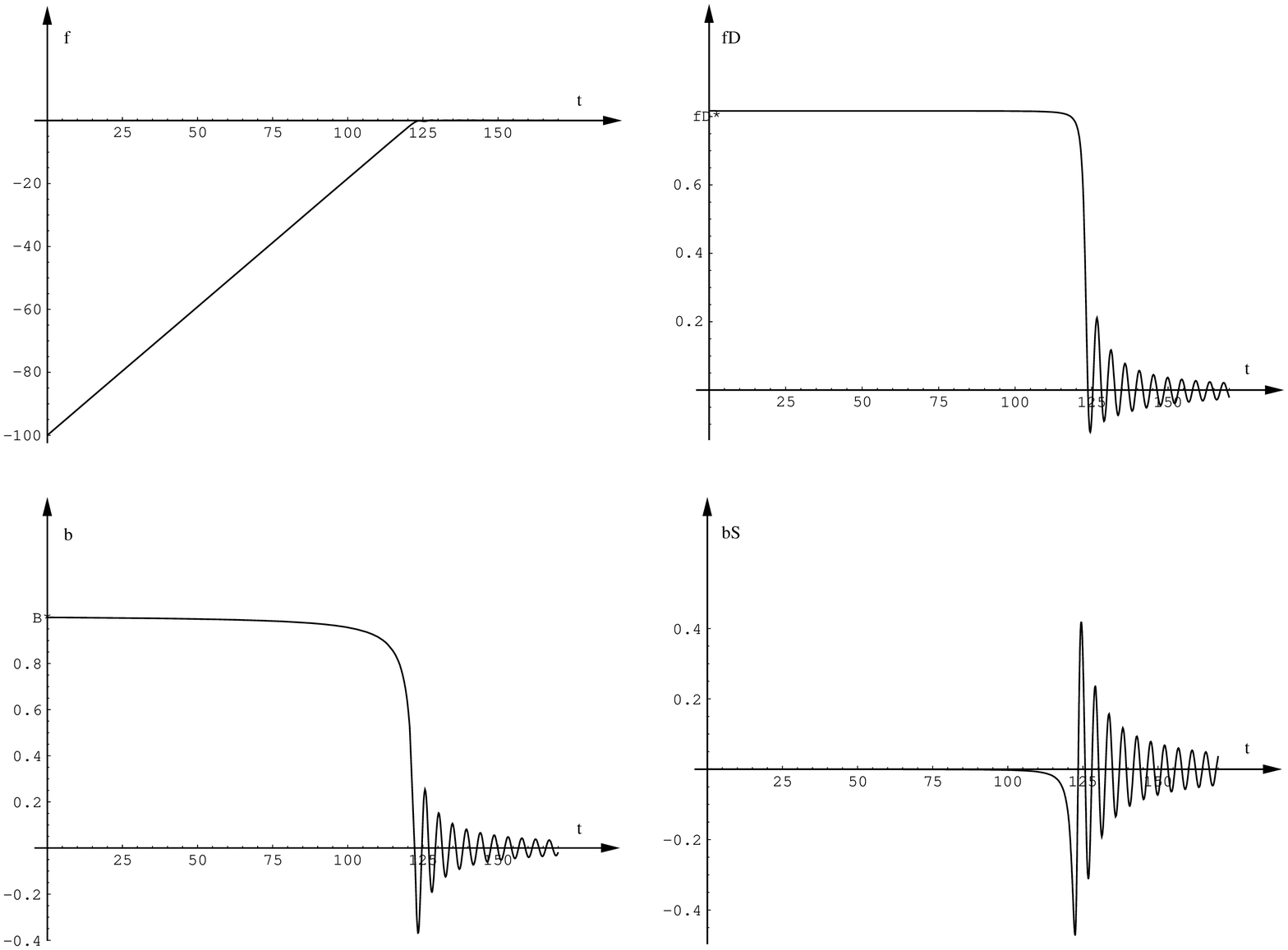}
\caption{\label{fi_dil} Dynamics of the system with $K(\phi)=-(\phi^2+0.5)^{-1}$, $Q=B^2/2$.
Initial values are $\phi=-100$, $\dot\phi=\dot \phi_*=\sqrt{2/3}$, 
$B=1$, $\dot B=0$}
\end{center}
\end{figure}

\begin{figure}[t]
\begin{center}
\psfrag{t}[tr]{\small$t$}
\psfrag{p}[bl]{$p$}
\psfrag{E}[bl]{$\varepsilon$}
\psfrag{w}[bl]{\small$w$}
\includegraphics[width=3in]{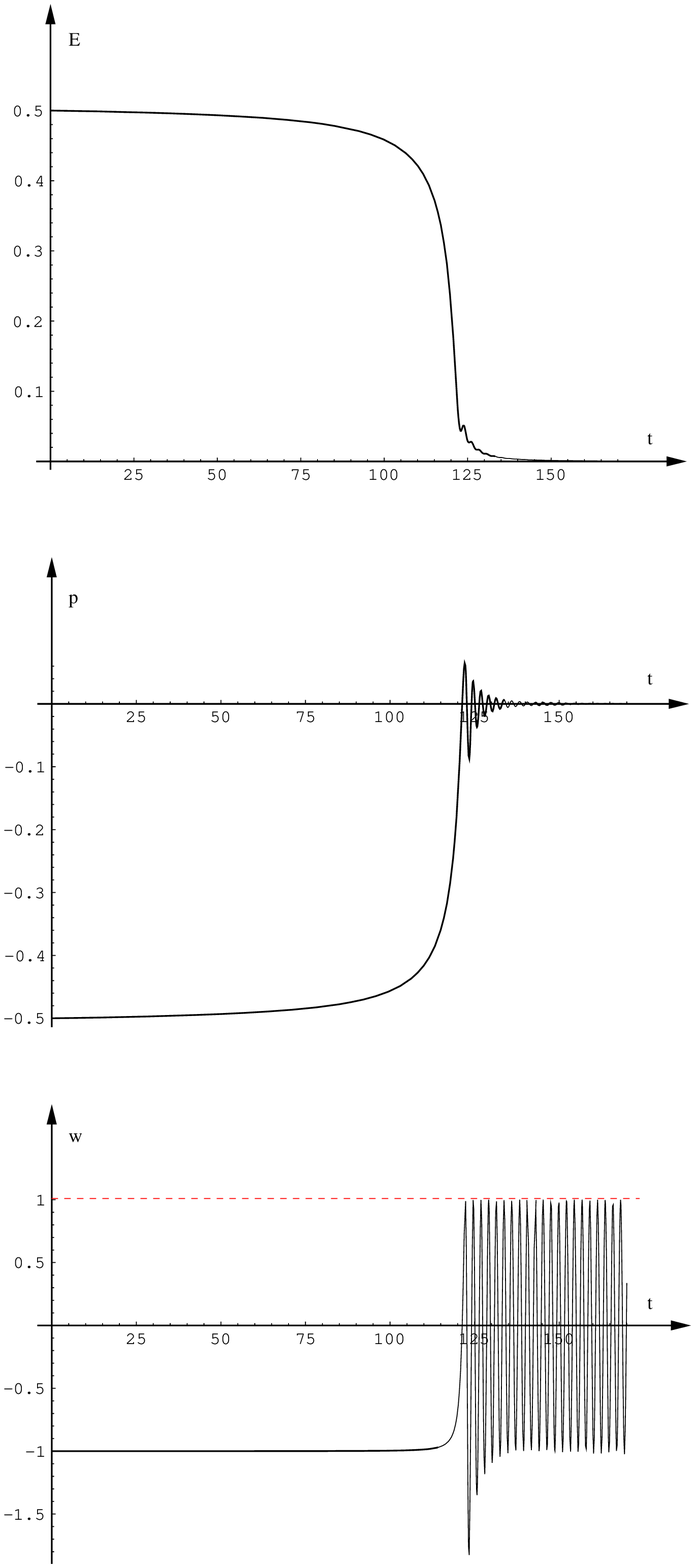}
\caption{\label{pew_dil} Energy density, pressure and equation of state 
for the system $K(\phi)=-(\phi^2+0.5)^{-1}$, $Q=B^2/2$.
Initial values are $\phi=-100$, $\dot\phi=\dot \phi_*=\sqrt{2/3}$, 
$B=1$, $\dot B=0$} 
\end{center}
\end{figure}

As the second example, we take 
\begin{equation}
  \label{K2}
  K(\phi)=-\frac{1}{\phi^2+\alpha^2},
\end{equation}
where $\alpha$ is some constant. There are several differences in the dynamics 
in comparison with the previous example. First of all, for the
choice (\ref{K2}) the inflationary regime is recovered for $\phi\to+\infty$
as well as for $\phi\to-\infty$. Another difference is that the exit 
in this example does not necessarily happen, because for sufficiently large $B_*$
the first term in (\ref{action}) is not ``strong'' enough to drive the system out 
of inflation. The maximum value of $B_*$ that still allows an exit can be 
estimated using Eq.~(\ref{Hsimm}), $B_*^{max}\sim \alpha^{-1}$.
The evolution of a system for $B_*<B_*^{max}$ is shown in Figs.~\ref{fi_dil} and
\ref{pew_dil}.  In addition the exit from inflationary stage in this case  
is milder than in the previous example. This is because the
function (\ref{K2}) is more widely spread than the function (\ref{K1}).
Asymptotically we have $\phi_e\to 0$ as $t\to\infty$, so the system 
arrives to the solution~(\ref{wkb}) with $m\simeq 1/\alpha$.

\section{Summary}

In this work we have introduced a new inflationary scenario where inflation is driven by 
a scalar field without a potential. We have shown that a very simple
Lagrangian, which is a function of the second derivatives only, 
naturally yields an inflationary solution. The initial conditions
which lead to inflation are generic. The inflationary
solution is manifestly stable with respect to the cosmological perturbations. 

We have proposed a mechanism of the graceful exit from the inflationary regime
into a dust-like state with $w\simeq 0$. In order to generate radiation, 
the usual mechanism of reheating similar to that in the ``standard''  
inflationary models can be applied to
our model. The detailed analysis of reheating, cosmological perturbations and 
observational consequences 
will be given elsewhere \cite{pert}.

\section*{Acknowledgements}

We are grateful to Viatcheslav Mukhanov for many illuminating discussions.
We also thank Sergei Winitzki for a number of valuable comments on the manuscript.
The work of E.B. was supported by the Deut\-sche
For\-schungs\-ge\-mein\-schaft within the Emmy Noether program and
the Russian Foundation for Basic Research, grant 04-02-16757-a.

\section*{References} 

\end{document}